\documentclass[10pt,conference]{IEEEtran}

\IEEEoverridecommandlockouts
\def\BibTeX{{\rm B\kern-.05em{\sc i\kern-.025em b}\kern-.08em
		T\kern-.1667em\lower.7ex\hbox{E}\kern-.125emX}}

\usepackage{dblfloatfix}    % To enable figures at the bottom of page

\usepackage{cite}
\usepackage{amsmath,amssymb,amsfonts}
\usepackage{bbm}

\usepackage{graphicx}
\usepackage{textcomp}
\usepackage{xcolor}
\usepackage{colortbl} % to use \rawcolor

\usepackage{url} %To make URL appears in bibliography

\usepackage{bm}
\usepackage{supertabular}
\usepackage{amsthm}
\theoremstyle{definition}

\usepackage[linesnumbered,ruled,vlined]{algorithm2e}% to use algorithm
\usepackage{pbox}
\usepackage{float} %to make tables and figures appear at specified location
\usepackage{subcaption} % to insert many plot in one figure
\usepackage{boldline,multirow}
\usepackage{array}

\usepackage{pbox}

\usepackage{dblfloatfix}

\usepackage[normalem]{ulem}

\usepackage{cancel} % to use goes to zero (or any constant) for an expression in math mode, usage: \cancelto{〈value〉}{〈expression〉}
\begin{document}

% To decrease space above and below equations
\setlength{\abovedisplayskip}{2pt}
\setlength{\belowdisplayskip}{2pt}

\title{POMDP-based Handoffs for User-Centric Cell-Free MIMO Networks}
\author{Hussein~A.~Ammar\IEEEauthorrefmark{1}, Raviraj~Adve\IEEEauthorrefmark{1}, Shahram~Shahbazpanahi\IEEEauthorrefmark{2}\IEEEauthorrefmark{1}, Gary~Boudreau\IEEEauthorrefmark{3}, and~Kothapalli~Venkata~Srinivas\IEEEauthorrefmark{3}% <-this % stops a space
	\\
	\IEEEauthorrefmark{1}University of Toronto, Dep. of Elec. and Comp. Eng., Toronto, Canada\\
	\IEEEauthorrefmark{2}University of Ontario Institute of Technology, Dep. of Elec. and Comp. Eng., Oshawa, Canada\\
	\IEEEauthorrefmark{3}Ericsson Canada, Ottawa, Canada\\
}

%\institute{University of Toronto\\
%\email{hussein.ammar@mail.utoronto.ca}}

\maketitle % Commented by Mr. H.A.A.

%%to make equation numbers size fixed and does not change when reducing size of equation
%\makeatletter
%\def\tagform@#1{\maketag@@@{\normalsize(#1)\@@italiccorr}}
%\makeatother

%%%%%%%%%%%%%%%%%%%%%%%%%%%%%%%%%%%%
% Performance
%%%%%%%%%%%%%%%%%%%%%%%%%%%%%%%%%%%%
\newcommand*{\nOfHOsTtrigLower}{$47\%$}
\newcommand*{\nOfHOsThreshtrig}{$70\%$}
\newcommand*{\CDFTtrigLower}{$82\%$}%{$89\%$}
\newcommand*{\CDFThreshtrig}{$90\%$}%{$97\%$}
%%%%%%%%%%%%%%%%%%%%%%%%%%%%%%%%%%%%

\vspace{-3em}
\begin{abstract}
We propose to control handoffs (HOs) in user-centric cell-free massive MIMO networks through a partially observable Markov decision process (POMDP) with the state space representing the discrete versions of the large-scale fading (LSF) and the action space representing the association decisions of the user with the access points. Our proposed formulation accounts for the temporal evolution and the partial observability of the channel states. This allows us to consider future rewards when performing HO decisions, and hence obtain a robust HO policy. To alleviate the high complexity of solving our POMDP, we follow a divide-and-conquer approach by breaking down the POMDP formulation into sub-problems, each solved individually. Then, the policy and the candidate cluster of access points for the best solved sub-problem is used to perform HOs within a specific time horizon. We control the number of HOs by determining when to use the HO policy. %and when should the user stick to the same serving cluster. 
Our simulation results show that our proposed solution reduces HOs by \nOfHOsTtrigLower\ compared to time-triggered LSF-based HOs and by \nOfHOsThreshtrig\ compared to data rate threshold-triggered LSF-based HOs. This amount can be further reduced through increasing the time horizon of the POMDP.% which also determines when the policy is deemed expired.
\end{abstract}

% \fi
\IEEEpeerreviewmaketitle

\section{Introduction}
Users in user-centric cell-free massive multiple-input multiple-output (UC-mMIMO) networks are served by neighboring access points or distributed units (DUs)~\cite{ammar9519163, ammar9570126}. This association produces an approximate uniform coverage for the whole network~\cite{interdonato2019ubiquitous}. However, with this scheme, moving users need to update their serving clusters very often, which could affect performance~\cite{ammar9650567}. A good strategy in managing handoffs (HOs) is to consider the temporal evolution of the channel states through some transition probabilities, which allows to connect the user to the best cluster of DUs based on current and future rewards.

Motivated by this, we propose to control HO decisions using a partially observable Markov decision process (POMDP) with the state space representing the discrete versions of the large-scale fading (LSF) and action space representing the association decisions of the user with the access points. A POMDP is a Markov process with feedback control and partially observable states~\cite{krishnamurthy2016partially, shani2013survey}. We follow a divide-and-conquer approach in which the POMDP problem is broken down into many sub-problems each with a candidate set of DUs. Then, the sub-problem that produces the best expected reward is selected with its policy to be used for performing HOs during a specific time window (time horizon). Our POMDP formulation accounts for the temporal statistics and the non-observability of the channel state information (CSI) for non-connected DUs to the user. Both of these features provide an advantage for POMDP-based HOs compared to LSF-based HOs. Our POMDP solution demonstrates efficiency when used in a framework for controlling the number of HOs while providing minimum data rate guarantees.

The use of a POMDP to control HOs in UC-mMIMO networks is novel and never studied in the literature. Existing studies focus on the impact of channel aging on the performance. Most studies consider static network environments, where the temporal evolution of the network is not considered. However, studying HOs requires implementing the concept of temporal evolution for the status of the network.

The studies in~\cite{CSIAging6608213, channelAgingMassiveMIMO8122014} are concerned with modeling and studying channel aging incurred from user movement. Additionally, the study in~\cite{9416909} investigates the impact of channel aging on the performance of cell-free massive MIMO networks by deriving closed-form expressions for the achievable rate in the uplink and downlink using coherent and noncoherent transmissions. Similarly, the study in~\cite{ChAgingPhaseNoise9471851} considers user mobility while focusing on phase noise resulting from imperfect synchronization of local oscillators at the DUs. The authors in~\cite{mobilityMMwaveCellFree9616361} partially consider controlling the number of HOs when using millimeter wave communication inside UC-mMIMO networks through a heuristic approach. However, the study does not focus on minimizing the number of HOs and the temporal aspects of the problem are not studied. All in all, the literature never proposes a tool with detailed components to model the HO problem, hence the importance of our work.

%The usage of POMDP allows to account for the temporal statistics to control handoff (HO), it also takes into consideration the partial observation of the states of the problem. 

%We model our problem as a POMDP, which is basically a Markov process with feedback control and a partial observability for the states of the problem. A POMDP is a generalization for a fully observable MDP. Hence, our usage of POMDP provides more freedom and less overhead in the taking our handoff decision compared to using MDP. The benefit is that we do not need to observe the whole states of the network channels to take the handoff decisions.

\section{Handoffs as a POMDP}\label{section:Model}
We consider the downlink of a network of DUs represented through the set $\mathcal{B}$. Each DU is equipped with $M$ antennas. We assume that the users are represented through the set $\mathcal{U}$, and they are moving inside the network. The users are served by the neighboring DUs using the UC-mMIMO scheme. The mobility of the users necessitates that each user $u$ updates its serving cluster, which is represented at decision cycle $t$ through the set $\mathcal{C}_u^{(t)}$.

Informed HO decisions require us to consider the temporal events experienced by the user and their effect on the performance. Based on this, we formulate the HO problem as a POMDP aiming to obtain a HO policy that instructs the user to which DUs to connect based on the user's movement and the status of the channels. Particularly, we wish to minimize the number of HOs as the user moves in the UC-mMIMO network, while maintaining a specific performance threshold.

\subsection{POMDP Formulation}
We consider a typical user moving through discrete time steps represented by decision cycles. HO can be initiated at the beginning of each decision cycle $t$. The channels between the user and the DUs constitute finite-state Markov processes over time. Furthermore, only the CSI between the user and the serving DUs (the user is currently connected to) can be known, which means the channels in the network can only be partially observed. This makes a POMDP a logical choice to model the HO operations. A POMDP contains a feedback control procedure that allows to determine the HO decisions based on the channels status and the system dynamics.

We define our POMDP framework using the following tuple
\begin{align}\label{eq:POMDP_model}
	& \mathcal{P}(\mathcal{B}_{\rm cand}, T_{\rm H})
	= \Big(\mathcal{S}(|\mathcal{B}_{\rm cand}|), \mathcal{A}(|\mathcal{B}_{\rm cand}|), \Omega(|\mathcal{B}_{\rm cand}|)
	\nonumber \\
	& \quad
	, {\bf P}_{\rm s, \mathcal{B}_{\rm cand}}^{(t)}%(\widetilde{\bf a})
	, {\bf P}_{\rm o, \mathcal{B}_{\rm cand}}^{(t)},
	{\bm \omega}^{(0)}_{\mathcal{B}_{\rm cand}}, R_{\mathcal{B}_{\rm cand}}\left({\bf s}^{(t)}, {\bf a}^{(t)}\right), T_{\rm H} \Big)
	,
\end{align}
where $\mathcal{S}(|\mathcal{B}_{\rm cand}|)$ is the state space when considering a set of candidate DUs $\mathcal{B}_{\rm cand} \subseteq \mathcal{B}$, $\mathcal{A}(|\mathcal{B}_{\rm cand}|)$ is the action space, $\Omega(|\mathcal{B}_{\rm cand}|)$ is the observation space, ${\bf P}_{\rm s, {\mathcal{B}_{\rm cand}}}^{(t)}$ is the transition probability matrix of the states, ${\bf P}_{\rm o, \mathcal{B}_{\rm cand}}^{(t)}$ is the observation probability matrix% for \emph{any} specific observation vector $\widetilde{\bf o}_l$
, ${\bm \omega}^{(0)}_{\mathcal{B}_{\rm cand}}$ is the initial belief distribution, $R_{\mathcal{B}_{\rm cand}}\left({\bf s}^{(t)}, {\bf a}^{(t)}\right)$ is the reward function which depends on the state ${\bf s}^{(t)}$ and action ${\bf a}^{(t)}$ found at decision cycle $t$, and $T_{\rm H}$ is the time horizon considered when solving the POMDP.

We use the indices $i$, $j$ and $l$ to refer to a specific state vector $\widetilde{\bf s}_i \in \mathcal{S}$, action vector $\widetilde{\bf a}_j \in \mathcal{A}$ and observation vector $\widetilde{\bf o}_l \in \Omega$, respectively. Also, %the notations 
$\mathcal{S}(|\mathcal{B}_{\rm cand}|), \mathcal{A}(|\mathcal{B}_{\rm cand}|), \Omega(|\mathcal{B}_{\rm cand}|)$ and $\mathcal{S}, \mathcal{A}, \Omega$ are used interchangeably.

\subsection{Components of the POMDP}
In this subsection, we define the components of the POMDP.
\subsubsection{State space}
$\mathcal{S}(|\mathcal{B}_{\rm cand}|) = \{\widetilde{\bf s}_1, \widetilde{\bf s}_2, \dots\}$, where the vector $\widetilde{\bf s}_i = [\widetilde{s}_{i, b_1 u}\ \widetilde{s}_{i, b_2 u} \dots]^T \in \mathbb{R}^{|\mathcal{B}_{\rm cand}| \times 1}$ represents a \emph{specific value} for the \emph{state} and $\widetilde{s}_{i, b_1 u}$ is a \emph{specific channel state} between DU $b_1 \in \mathcal{B}_{\rm cand}$ and typical user $u$.

We define the \emph{channel state} as the discrete version of the LSF, where the values of LSF are quantized into $2$ levels based on a threshold $\beta_{\rm threshold}$ (two-state channel). Hence, we classify the channel state $s_{bu}^{(t)}$ at decision cycle $t$ between DU $b \in \mathcal{B}_{\rm cand}$ and the typical user as either a good channel $\widetilde{\beta}_{1}$ if $\beta_{bu}^{(t)} > \beta_{\rm threshold}$ where $\beta_{bu}^{(t)}$ is the LSF between DU $b \in \mathcal{B}_{\rm cand}$ and the typical user at a decision cycle $t$, or a bad one $\widetilde{\beta}_{0}$ otherwise.%as shown below
%\begin{align}\label{eq:twoChannelState}
%	s_{bu}^{(t)}
%	=
%	\begin{cases}
%		\widetilde{\beta}_{1}, & \text{if}\ \beta_{bu}^{(t)} > \beta_{\rm threshold},
%		\\
%		\widetilde{\beta}_{0}, & \text{if}\ \beta_{bu}^{(t)} \le \beta_{\rm threshold},
%	\end{cases}
%\end{align}
%where $\beta_{bu}^{(t)}$ is the large-scale fading between DU $b \in \mathcal{B}_{\rm cand}$ and the typical user at a decision cycle $t$.

As a result, we use the vector ${\bf s}^{(t)} = [s_{b_1u}^{(t)}\ s_{b_2u}^{(t)} \dots]^T \in \mathbb{R}^{|\mathcal{B}_{\rm cand}| \times 1}$ to denote the state at decision cycle $t$, which is a random variable. To prevent confusion, ${\bf s}^{(t)}$ simply represents the \emph{state} at decision cycle $t$, while $s_{bu}^{(t)}$ represents a single \emph{channel state}.% If we have $2$ possible values for $s_{bu}^{(t)}$, then we end up having $2^{|\mathcal{B}_{\rm cand}|}$ possible combinations for the states.

\subsubsection{Action space}
$\mathcal{A}(|\mathcal{B}_{\rm cand}|) = \{ \widetilde{\bf a}_1, \widetilde{\bf a}_2, \dots \}$, where the vector $\widetilde{\bf a}_j = [ \widetilde{a}_{j,b_1 u}\ \widetilde{a}_{j,b_2 u} \dots ]^T \in \mathbb{B}^{|\mathcal{B}_{\rm cand}| \times 1}$ represents a specific action for the POMDP, with $\widetilde{a}_{j,b u}$ being the action between DU $b \in \mathcal{B}_{\rm cand}$ and the user.% Similar to the state space, the action space is all the possible combinations of the different actions.

At a specific $t$, user $u$ can either connect to DU $b \in \mathcal{B}_{\rm cand}$ or not. We use $a_{bu}^{(t)} \in \{0, 1\}$ to represent the decision of either connect ($a_{bu}^{(t)} = 1$) or disconnect ($a_{bu}^{(t)} = 0$) to DU $b$. We use the notation ${\bf a}^{(t)} = [a_{b_1u}^{(t)}\ a_{b_2u}^{(t)}\ \dots ]^T \in \mathbb{B}^{|\mathcal{B}_{\rm cand}| \times 1}$ to denote the action for user $u$ at $t$ with the DUs $\mathcal{B}_{\rm cand}$, i.e., the vector ${\bf a}^{(t)}$ represents the action decided at $t$ by the POMDP policy. 

%It represents all the possible combinations of the different actions.  

We further assume that a user is connected to $B_{\rm con} < |\mathcal{B}_{\rm cand}|$ DUs. This is represented through the condition $\sum_{b \in \mathcal{B}_{\rm cand}} a_{bu}^{(t)} = B_{\rm con}$. Using a fixed $B_{\rm con}$ allows us to shrink the action space compared to an unconstrained number of associations. Based on this, $\mathcal{A}(|\mathcal{B}_{\rm cand}|)$ contains ${|\mathcal{B}_{\rm cand}| \choose B_{\rm con} } = \frac{|\mathcal{B}_{\rm cand}|!}{B_{\rm con}! \left(|\mathcal{B}_{\rm cand}| - B_{\rm con} \right)!}$ %$2^{(|\mathcal{B}| - B_{\rm con}) U}$ 
possible combinations of actions.% Please note that the size of the action space can be very huge if we consider that the user can connect to a random number of DUs, hence, placing this constraint is crucial to manage the complexity of the model.

\subsubsection{Transition probability} We use ${\bf P}_{\rm s, {\mathcal{B}_{\rm cand}}}^{(t)} \in \mathbb{R}^{|\mathcal{S}| \times |\mathcal{S}|}$ to denote the transition probability matrix for the states at decision cycle $t$. For $\tilde{\bf s}_i, \tilde{\bf s}_{i'} \in \mathcal{S}(|\mathcal{B}_{\rm cand}|)$, element $[{\bf P}_{\rm s, {\mathcal{B}_{\rm cand}}}^{(t)}]_{i,i'} \in [0,1]$ (at row $i$ and column $i'$ of the matrix) can be defined as
\begin{align}\label{eq:Transition_prob}
	%	[{\bf P}_{\rm s}(\widetilde{\bf a}_l)]_{j,j'}
	%	=
	%	\mathbb{P}({\bf s}^{(t+1)} = \widetilde{\bf s}_{j'}|\ {\bf s}^{(t)} = \widetilde{\bf s}_j, {\bf a}^{(t)} = \widetilde{\bf a}_l), \quad \check{\bf s}_j, \check{\bf s}_{j'} \in \mathcal{S}
	%
	[{\bf P}_{\rm s, {\mathcal{B}_{\rm cand}}}^{(t)}]_{i,i'}
	&=
	\mathbb{P}({\bf s}^{(t)} = \widetilde{\bf s}_{i'}|\ {\bf s}^{(t-1)} = \widetilde{\bf s}_i)
	\nonumber \\
	&=
	\prod_{b \in \mathcal{B}_{\rm cand}}
	\mathbb{P}( s_{bu}^{(t)} = \widetilde{s}_{i',bu} |\ s_{bu}^{(t-1)} = \widetilde{s}_{i,bu} )
\end{align}
Each row in the matrix ${\bf P}_{\rm s, {\mathcal{B}_{\rm cand}}}^{(t)}$ sums up to $1$.

\begin{figure}[t]%[H]
	\centering
	\includegraphics[width=0.8\columnwidth]{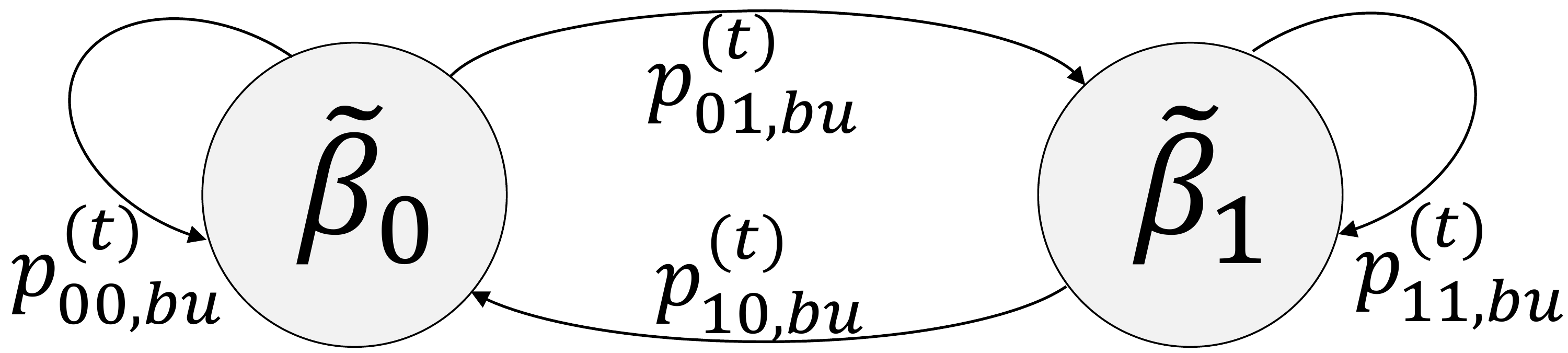}
%	\vspace{-0.5em}
	\caption{Transition probability diagram for the \emph{channel state} between DU $b$ and user at the beginning of decision cycle~$t$.} 
	\label{fig:mob_twoStateChannel}
	\vspace{-1.5em}
\end{figure}

Based on the two-state channel, we end up with the transition probabilities of the channel states depicted in Fig.~\ref{fig:mob_twoStateChannel}. To characterize these probabilities we only need to characterize $p_{11}^{(t)}$ and $p_{01}^{(t)}$, because $p_{11}^{(t)} + p_{10}^{(t)} = 1$ and $p_{00}^{(t)} + p_{01}^{(t)} = 1$.

Hence, each term inside the product in~\eqref{eq:Transition_prob} is defined as 
\begin{align}
	&\mathbb{P}( s_{bu}^{(t)} = \widetilde{s}_{i',bu} |\ s_{bu}^{(t-1)} = \widetilde{s}_{i,bu} )
	=
	\nonumber \\
	&
	\begin{cases}
		p_{11, bu}^{(t)},& \text{if}\ \widetilde{s}_{i',bu} = \widetilde{\beta}_{1}, \widetilde{s}_{i,bu} = \widetilde{\beta}_{1} \\
		p_{01, bu}^{(t)},& \text{if}\ \widetilde{s}_{i',bu} = \widetilde{\beta}_{1}, \widetilde{s}_{i,bu} = \widetilde{\beta}_{0}\\
		1 - p_{11, bu}^{(t)},& \text{if}\ \widetilde{s}_{i',bu} = \widetilde{\beta}_{0}, \widetilde{s}_{i,bu} = \widetilde{\beta}_{1}\\
		1 - p_{01, bu}^{(t)},& \text{if}\ \widetilde{s}_{i',bu} = \widetilde{\beta}_{0}, \widetilde{s}_{i,bu} = \widetilde{\beta}_{0}
	\end{cases}
\end{align}

The transition probabilities $p_{11, bu}^{(t)}$ and $p_{01, bu}^{(t)}$ for the channel state between DU $b \in \mathcal{B}_{\rm cand}$ and the user $u$ can be calculated based on the user mobility through the correlated shadowing model between two different locations of the user before and after mobility~\cite{CorrelatedShadowing4357088, CorrelatedShadowing104090}. Check~\cite{POMDP_J} for more details.

\subsubsection{Observation Space $\Omega(|\mathcal{B}_{\rm cand}|)$} The vector ${\bf o}^{(t)} = [o_{b_1u}^{(t)} \dots ]^T \in \mathbb{R}^{|\mathcal{B}_{\rm cand}| \times 1}$ is used to represent the observation at decision cycle $t$. We assume that the channel state between the user and the \emph{currently connected} DUs is known, i.e., if $a_{bu}^{(t)} = 1$, then $o_{bu}^{(t)} = s_{bu}^{(t)}$ and the channel state is observable. If the user is not connected to DU $b$, then the channel state $s_{bu}^{(t)}$ is not observable and this will affect the observability of the state ${\bf s}^{(t)}$. In such a case, a probabilistic approach should be constructed through the observation distribution.% which we is done through the observation distribution. 
%${\bf o}^{(t)} = \left( {\rm diag}\left( {\bf a}^{(t)} \right) {\bf s}^{(t)} \right)_{{\bf a}^{(t)}\ne 0} \in \mathbb{R}^{B_{\rm con}}$
%, where $(\cdot)_{{\bf a}^{(t)}\ne0}$ is a selection operator that selects the nonzero elements based on the vector ${\bf a}^{(t)}$.

\subsubsection{Observation distribution} We use ${\bf P}_{\rm o, \mathcal{B}_{\rm cand}}^{(t)} \in \mathbb{R}^{|\mathcal{S}|  \times |\Omega| \times |\mathcal{A}|}$ as the observation probability matrix. %for a specific observation vector $\widetilde{\bf o}_l$. 
Herein, we assume that we can only observe the channel states toward the DUs that the user is currently connected to. For $\widetilde{\bf s}_i \in \mathcal{S}(|\mathcal{B}_{\rm cand}|)$, $\widetilde{\bf o}_l \in \Omega(|\mathcal{B}_{\rm cand}|)$, $\widetilde{\bf a}_j \in \mathcal{A}(|\mathcal{B}_{\rm cand}|)$, element $\left[{\bf P}_{\rm o, \mathcal{B}_{\rm cand}}^{(t)}(\widetilde{\bf o}_l)\right]_{i, l,j} \in [0,1]$ is defined as
\begin{align}\label{eq:observ_prob}
	&\left[{\bf P}_{\rm o, \mathcal{B}_{\rm cand}}^{(t)}\right]_{i,l,j}
	=
	\mathbb{P}({\bf o}^{(t)} = \widetilde{\bf o}_l|\ {\bf s}^{(t)} = \widetilde{\bf s}_i, {\bf a}^{(t)} = \widetilde{\bf a}_j)\nonumber \\
	& \quad
	=
	\prod_{b \in \mathcal{B}_{\rm cand}} \mathbb{P}\left( {o}_{bu}^{(t)} = \widetilde{o}_{l,bu} |\  {\bf s}^{(t)} = \widetilde{\bf s}_i, {\bf a}^{(t)} = \widetilde{\bf a}_j\right)
	\nonumber \\
	& \quad
	=
	\Bigg(
	\prod_{b \in \mathcal{C}_u^{(t)}}
		\zeta_{bu, ilj}
	\Bigg)
	\prod_{b \in \mathcal{B}_{\rm cand} \backslash\mathcal{C}_u^{(t)}} 
		\bar{\zeta}_{bu, ilj}
\end{align}
where $\zeta_{bu, ilj}$ and $\bar{\zeta}_{bu, ilj}$ correspond to the observed and non-observed channel state, respectively, and they are defined as
\begin{align}
\zeta_{bu, ilj} &= 
\mathbb{P}\left( {o}_{bu}^{(t)} = \widetilde{s}_{l,bu} |\ {\bf s}^{(t)} = \widetilde{\bf s}_i, {\bf a}^{(t)} = \widetilde{\bf a}_j\right)
\nonumber \\
&=
\begin{cases}
	1, & \text{if}\ l = i \\
	0, & \text{otherwise}
\end{cases}	
\\
\bar{\zeta}_{bu, ilj} &= \ 
\mathbb{P}\left( {o}_{bu}^{(t)} = \widetilde{o}_{l,bu} |\  {\bf s}^{(t)} = \widetilde{\bf s}_i, {\bf a}^{(t)} = \widetilde{\bf a}_j\right)
\nonumber \\
&=
\begin{cases}
	\bar{p}_{1, bu}^{(t)}, & \text{if}\ \widetilde{o}_{l,bu} = \widetilde{\beta}_{1} \\
	\bar{p}_{0, bu}^{(t)}, & \text{if}\ \widetilde{o}_{l,bu} = \widetilde{\beta}_{0} 
\end{cases}
\label{eq:observProb_unknown}
\end{align}
We note that this leads to $\sum_{i = 1}^{|\mathcal{S}|} \left[{\bf P}_{\rm o, \mathcal{B}_{\rm cand}}^{(t)}\right]_{i,l,j} = 1$ for any $l$ and $j$, which is a requirement for the observation probability matrix. The probabilities $\bar{p}_{1, bu}^{(t)}$ and $\bar{p}_{0, bu}^{(t)}$ are calculated as
\begin{align}\label{eq:p1}
	& \bar{p}_{1, bu}^{(t)}
	\triangleq
	\mathbb{P}(s_{bu}^{(t)} = \widetilde{\beta}_{1} )
	%	\nonumber \\
	%	&
	\stackrel{(a)}
	{=}
	\mathbb{P}(\beta_{bu}^{(t)} > \beta_{\rm threshold} )
%	\nonumber \\
%	&
%	\stackrel{(b)}
%	{=}
%	\resizebox{0.61\columnwidth}{!}
%	{$
%	\mathbb{P}\left(\bar{\kappa}_{bu}^{(t)} > \frac{10}{\sigma_\mathrm{sh|dB}}\log_{10}\left(\frac{\beta_{\rm threshold}}{{\rm PL}^{(t)}}\right) \right)
%	$}
%	%	\nonumber \\
%	%	&
%	=
%	{\rm Q}\left( \dot{k}^{(t)} \right)
%	,
\end{align}
where $(a)$ follows from the two-state channel, also
%, $(b)$ follows from the correlated shadowing model, %~\eqref{eq:shad_model} for $t=0$, 
%and $\bar{\kappa}_{bu}^{(t)} \sim \mathcal{N}(0, 1)$. Furthermore, by definition, the probability of observing $\widetilde{\beta}_{0}$ between DU $b$ and the typical user equals
\begin{align}\label{eq:p0}
	\bar{p}_{0, bu}^{(t)} 
	\triangleq
	1 - \bar{p}_{1, bu}^{(t)}
\end{align}
%If we have an MDP (fully observable), the observation probability would be represented by an identity matrix.

\subsubsection{Reward} $R_{\mathcal{B}_{\rm cand}}({\bf s}^{(t)}, {\bf a}^{(t)})$ is the reward when executing action ${\bf a}^{(t)}$ at state ${\bf s}^{(t)}$. This function is based on the achievable rate, and it is defined in the next section after we define some important system dynamics.

\subsubsection{Time horizon $T_{\rm H}$} It is the number of decision cycles (in the future) to consider when optimizing the cumulative expected discounted
reward of the POMDP.% The time horizon allows us to consider future rewards when taking our actions.

\subsubsection{Belief} The belief defines the posterior probability distribution over the state space and represents the knowledge of the POMDP decision maker about the state based on past actions and observations. %Since the states are not fully observable, the agent has to consider a complete history of its past actions and observations to determine the current action. 
We use ${\bm \omega}^{(t)}_{\mathcal{B}_{\rm cand}} = [\omega_{1, \mathcal{B}_{\rm cand}}^{(t)}, \dots, \omega_{|\mathcal{S}|, \mathcal{B}_{\rm cand}}^{(t)}]$ as the belief vector at decision cycle $t$, where $\omega_{i, \mathcal{B}_{\rm cand}}^{(t)}$ is the probability of ${\bf s}^{(t)}$ being equal to a particular value $\widetilde{\bf s}_i \in \mathcal{S}$, given all the action and observation history from $t=0$ till $(t-1)$. Hence, 
\begin{align}\label{eq:belief_w}
	&\omega_{i, \mathcal{B}_{\rm cand}}^{(t)} = \mathbb{P}\left( {\bf s}^{(t)} = \widetilde{\bf s}_i |\ \mathcal{H}_{t-1}\right)
	\nonumber \\
	&=
	\prod_{b \in \mathcal{B}_{\rm cand}}
	\mathbb{P}\left( s_{bu}^{(t)} = \widetilde{s}_{i,bu}  |\ \mathcal{H}_{t-1}\right)
	,\quad i = 1,\dots, |\mathcal{S}|
\end{align}
where $\mathcal{H}_{t-1} = \{ {\bf o}^{(t-1)}, {\bf a}^{(t-1)}, \mathcal{H}_{t-2}\}$ denotes the history of observations and actions.

We define the belief of state $s_{bu}^{(t)}$ being $\widetilde{\beta}_{1}$ as $\Upsilon_{bu}^{(t)}$ using
\begin{align}\label{eq:ProbGoodStateElement}
&\Upsilon_{bu}^{(t)}
\triangleq
\mathbb{P}\left( s_{bu}^{(t)} = \widetilde{\beta}_{1}  |\ \mathcal{H}_{t-1}\right)
\nonumber \\
&=
\begin{cases}
	p_{11, bu}^{(t)},& \text{if}\ a_{bu}^{(t-1)} = 1, s_{bu}^{(t-1)} = \widetilde{\beta}_{1}, \\
	p_{01, bu}^{(t)},&  \text{if}\ a_{bu}^{(t-1)} = 1, s_{bu}^{(t-1)} = \widetilde{\beta}_{0},\\
	\begin{aligned}
	&\Upsilon_{bu}^{(t-1)} p_{11, bu}^{(t)}
	\\
	&
	+ \left(1 - \Upsilon_{bu}^{(t-1)}\right) p_{01, bu}^{(t)},
	\end{aligned}
	&  \text{if}\ a_{bu}^{(t-1)} = 0
\end{cases}
\end{align}
The third case term in~\eqref{eq:ProbGoodStateElement} is written in a probabilistic fashion because when $a_{bu}^{(t-1)} = 0$ (user is not connected to DU $b$) the channel state $s_{bu}^{(t-1)}$ is unobservable.

For the initial belief ${\bm \omega}^{(0)}_{\mathcal{B}_{\rm cand}}$, we make use of the initial transition probabilities of the channels to define the elements $\{\omega_{i, \mathcal{B}_{\rm cand}}^{(0)} : i = 1,\dots, |\mathcal{S}|\}$ of the initial belief as 
\begin{align}\label{eq:belief_w_initial}
\omega_{i, \mathcal{B}_{\rm cand}}^{(0)} &=
\mathbb{P}\left( {\bf s}^{(0)} = \widetilde{\bf s}_i |\ \mathcal{H}_{-1}\right)
=
\resizebox{0.43\columnwidth}{!}
{$
\prod_{b \in \mathcal{B}_{\rm cand}}
\mathbb{P}\left( s_{bu}^{(0)} = \widetilde{s}_{i,bu} \right)
$}
.
\end{align}
%\begin{align}
%&\text{with} \quad
%\mathbb{P}\left( s_{bu}^{(0)} = \widetilde{s}_{i,bu} \right)
%=
%\nonumber \\
%&
%\begin{cases}
%	1,& \text{if}\ b \in \mathcal{C}_u^{(0)}, \beta_{bu}^{(0)} > \beta_{\rm threshold}, \widetilde{s}_{i,bu} = \widetilde{\beta}_{1} \\
%	1,& \text{if}\ b \in \mathcal{C}_u^{(0)}, \beta_{bu}^{(0)} \le \beta_{\rm threshold}, \widetilde{s}_{i,bu} = \widetilde{\beta}_{0} \\
%	0,& \text{if}\ b \in \mathcal{C}_u^{(0)}, \beta_{bu}^{(0)} > \beta_{\rm threshold}, \widetilde{s}_{i,bu} = \widetilde{\beta}_{0} \\
%	0,& \text{if}\ b \in \mathcal{C}_u^{(0)}, \beta_{bu}^{(0)} \le \beta_{\rm threshold}, \widetilde{s}_{i,bu} = \widetilde{\beta}_{1} \\
%	\bar{p}_{1, bu}^{(0)},& \text{if}\ b \notin \mathcal{C}_u^{(0)}, \widetilde{s}_{i,bu} = \widetilde{\beta}_{1} \\
%	\bar{p}_{0, bu}^{(0)},&  \text{if}\ b \notin \mathcal{C}_u^{(0)}, \widetilde{s}_{i,bu} = \widetilde{\beta}_{0}
%\end{cases}
%\end{align}
%where the terms $\bar{p}_{1, bu}^{(t)}$ and $\bar{p}_{0, bu}^{(t)}$ are defined in~\eqref{eq:p1} and~\eqref{eq:p0}, respectively.

\subsection{Objective} 
The objective function is defined as the long term reward averaged over the different states as follows
\begin{align}\label{eq:obj_POMDP}
	J_{\Pi}({\bm \omega}^{(0)}_{\mathcal{B}_{\rm cand}})
	=
	\resizebox{0.64\columnwidth}{!}
	{$
	\mathbb{E}_{\mathcal{S}} \left\{ \sum_{t = 1}^{T_{\rm H}} \gamma^t R_{\mathcal{B}_{\rm cand}}\left({\bf s}^{(t)}, {\bf a}^{(t)}\right) | {\bm \omega}^{(0)}_{\mathcal{B}_{\rm cand}} \right\}
	$}
	,
\end{align}
where $0\le \gamma < 1$ is a discount factor for future rewards.% The summation over $t$ in~\eqref{eq:obj_POMDP} goes to $T_{\rm H}$.

Finally, the aim is to determine an optimal policy $\Pi^\star$ that maximizes the objective function in~\eqref{eq:obj_POMDP}, i.e.,
\begin{align}
	\Pi^\star = \operatorname{arg}\ \underset{\Pi}{\operatorname{max}} \quad J_{\Pi}({\bm \omega}^{(0)}_{\mathcal{B}_{\rm cand}}),\ \text{for\ any}\ {\bm \omega}^{(0)}_{\mathcal{B}_{\rm cand}}
\end{align}

\section{Problem Dynamics}
User movement causes the communication channel to evolve quickly with temporal correlation. Within a decision cycle $t$ (which contains many channel uses), channel change can be modeled through channel aging~\cite{channelAgingMassiveMIMO8122014} that is the mismatch between the CSI at the time the channel was estimated and at the time it was used for data transmission. In the following we will use ${\bf h}_{bu}[n] \triangleq \sqrt{ \beta_{bu} } {\bf g}_{bu}[n] \in \mathbb{C}^{M \times 1}$ as the channel between DU $b$ and the user at a specific channel use (instant $n$) within a decision cycle, where ${\bf g}_{bu}[n] \sim \mathcal{CN}\left({\bf 0}, {\bf I}_M \right)$ is the small-scale fading, and $\beta_{bu}$ is the large-scale fading that accounts for the shadowing and the path loss. Because the analysis is within a single decision cycle, we have dropped the superscript $(t)$.

\subsection{Channel Aging and Estimation Errors}
Channel aging is modeled using Jakes' model as follows~\cite{channelAgingMassiveMIMO8122014}:
\begin{align}\label{eq:channelEvolution}
	{\bf g}_{bu}[n] = \rho_{u}[n] {\bf g}_{bu}[0] + \bar{\rho}_{u}[n] {\bf v}_{bu}[n],
\end{align}
where ${\bf v}_{bu}[n] \sim \mathcal{CN}\left({\bf 0}, {\bf I}_M \right)$ is the independent innovation component at instant $n$, $\rho_{u}[n]$ is the temporal correlation coefficient of user $u$ between channel realizations at instants $0$ and $n$, with $0 \le \rho_{u}[n] \le 1$, and $\bar{\rho}_{u}[n] = \sqrt{ 1 - |\rho_{u}[n]|^2 }$. The temporal correlation is defined as~\cite{channelAgingMassiveMIMO8122014}
\begin{align}\label{eq:aging}
	\rho_{u}[n - n']
%	=
%	[{\bf C}_{bm,u}]_{n n'} = \mathbb{E}\left\{ g_{bm,u}[n] g_{bm,u}[n'] \right\} 
	\triangleq
	J_0 \left( 2 \pi \left( n - n' \right) f_{{\rm{D}}_u} T_{\rm s} \right)
\end{align}
where $J_0(\cdot)$ is the zeroth-order Bessel function of the first kind, $f_{{\rm{D}}_u}$ is the maximum Doppler shift, and $T_{\rm s}$ is the sampling period of each channel use.

Based on~\eqref{eq:channelEvolution}, during a pilot training phase, $\{{\bf g}_{bu}[n]: n < n_{\rm est}\}$ can be related to ${\bf g}_{bu}[n_{\rm est}]$ at time instant $n_{\rm est}$, where $n_{\rm est}$ is the time instant of channel estimation, as follows:
\begin{align}\label{eq:channelEvolutionChanEst}
	{\bf g}_{bu}[n] = \rho_{u}[n_{\rm est} - n] {\bf g}_{bu}[n_{\rm est}] + \bar{\rho}_{u}[n_{\rm est} - n] {\bf v}_{bu}[n]
\end{align}

Using linear minimum mean square error (LMMSE), DU $b$ estimates the channel of user $u \in \mathcal{U}_i$~as
\begin{align}\label{eq:est_chan}
	{\bf \hat{h}}_{bu}[n_{\rm est}]
	=
	\frac{ \rho_{u}[n_{\rm est} - i ] \sqrt{p^{({\rm u})}} \beta_{bu}}{ \sum_{u' \in \mathcal{U}_i} p^{({\rm u})} \beta_{bu'} + \sigma_{\rm z}^2 }
	{\bf y}_{b}[i],
	\quad \text{for}\ u \in \mathcal{U}_i
\end{align}
where $p^{({\rm u})}$ is the pilot uplink transmit power of user, ${\bf y}_{b}[i]$ is the signal received at DU $b$ at channel use instant $i$ during the uplink pilot training phase, the set $\mathcal{U}_i$ represents the users using the same pilot as user $u$, and $\sigma_{\rm z}^2$ is the variance of noise.

Using LMMSE estimation, the channel estimation error $\mathrm{ {\bf e}}_{bu}[n_{\rm est}] = {\bf h}_{bu}[n_{\rm est}] - {\bf \hat{h}}_{bu}[n_{\rm est}]$ is uncorrelated with the estimated channel ${\bf \hat{h}}_{bu}[n_{\rm est}]$ and is distributed as $\mathrm{ {\bf e}}_{bu}[n_{\rm est}] \sim \mathcal{CN}\left({\bf 0}, {\bm \Theta_{bu}}\right)$, where the ${\bm \Theta_{bu}} \triangleq \beta_{bu} {\bf I}_M - \psi_{bu} {\bf I}_M$, with
\begin{align}\label{eq:var_estCh}
	\psi_{bu}[n_{\rm est}]
	=
	\frac{\rho_{u}^2[ n_{\rm est} - i  ] p^{({\rm u})} \beta_{bu}^2 }{ \sum_{u' \in \mathcal{U}_i} p^{({\rm u})} \beta_{bu'} + \sigma_{\rm z}^2 },
	\quad \text{for}\ u \in \mathcal{U}_i
\end{align}

\subsection{Achievable Rate and Reward Function}
For mathematical tractability, and to focus on the problem, we assume that the DUs use conjugate beamforming to serve the users. Hence, each DU uses the channels estimated at time instant $n_{\rm est}$ to construct the beamformer to send the data to its users $\mathcal{E}_b$ within instants $n_{\rm est}$ till $\tau_{\rm c}$, where $\tau_{\rm c}$ is the number of channel uses per a single communication phase composed of pilot training and data transmission.

Using the channel aging model in~\eqref{eq:channelEvolutionChanEst} and the channel estimation error in~\eqref{eq:var_estCh}, we can characterize the performance through a lower bound for the ergodic achievable rate, where the signal received at the user can be quantized into a desired signal, channel aging, and beamformer uncertainity and noise terms. Then, we can define the reward function for the POMDP using the signal-to-noise (SNR) version of the achievable rate as
\begin{align}\label{eq:reward_SingleU}
	&R_{\mathcal{B}_{\rm cand}}\left({\bf s}^{(t)}, {\bf a}^{(t)}\right) =
	\nonumber \\
	& \quad
	\frac{1}{\tau_{\rm c}} \sum_{n=n_{\rm est}}^{\tau_{\rm c}} \log \left( 1 + \frac{ \xi_{1,{\rm S}, u}\left[n, {\bf s}^{(t)}, {\bf a}^{(t)}\right]
	}{\displaystyle
		\xi_{2,3,{\rm S}, u}\left[n, {\bf s}^{(t)}, {\bf a}^{(t)}\right] + \sigma_{\rm z}^2 } \right) 
\end{align}
where $\sigma_{\rm z}^2$ is the power of the noise and
{\allowdisplaybreaks
	\begin{align}
		&
		\resizebox{0.25\columnwidth}{!}
		{$
		\xi_{1,{\rm S}, u}\left[n, {\bf s}^{(t)}, {\bf a}^{(t)}\right]
		$}
		=
%		\nonumber \\
%		& \quad
	\resizebox{0.7\columnwidth}{!}
	{$
		M p^{({\rm d})}
		\rho_{u}^2[n - n_{\rm est}]
		\left|
		\sum_{b \in \mathcal{B}_{\rm cand}}
		a_{bu}^{(t)}
		\sqrt{
			\frac{\psi_{{\rm S}, bu} [s_{bu}^{(t)} ]}{|\mathcal{E}_b|}
		} 
		\right|^2
	$}
\label{eq:ds}
		\\
		&
		\xi_{2,3,{\rm S}, u}\left[n, {\bf s}^{(t)}, {\bf a}^{(t)}\right]
		=
		M
		p^{({\rm d})}
		\sum_{b \in \mathcal{B}_{\rm cand}} 
		\frac{a_{bu}^{(t)} s_{bu}^{(t)} }
		{
			|\mathcal{E}_b|
		}
		\label{eq:BU_CA}
		%%%%%%%%%%%%%%%%%%%%%
		%%%%%%%%%%%%%%%%%%%%%
		%%%%%%%%%%%%%%%%%%%%%
\end{align}}
$\!\!$where we made use of the notations to rewrite the variance of the estimated channel in~\eqref{eq:var_estCh} for a single-user case as
\begin{align}\label{eq:psi_POMDP_SingleU}
	\psi_{{\rm S}, bu}[s_{bu}^{(t)}]
	=
	\frac{\rho_{u}^2[ n_{\rm est} - i ] p^{({\rm u})} \big(s_{bu}^{(t)}\big)^2 }{ \sigma_{\rm z}^2 },
	\quad \text{for}\ u \in \mathcal{U}_i
	,
\end{align}
and the allocated power $\eta_{{\rm S}, bu}[n_{\rm est}]$ to user $u$ as
\begin{align}\label{eq:allocaPower}
	\eta_{{\rm S}, bu}[n_{\rm est}] =
	\frac{p^{({\rm d})}}
	{
		M |\mathcal{E}_b| \psi_{{\rm S}, bu}[s_{bu}^{(t)}]
	}.
\end{align}

Moreover, we used the fact that we can relate any term $c_{bu}$ that contains a summation over $\mathcal{C}_u^{(t)}$ to a summation over all the DUs $\mathcal{B}_{\rm cand}$ through
\begin{align}\label{eq:action_POMDP}
	\sum_{b \in \mathcal{C}_u^{(t)}} c_{bu}
	=
	\sum_{b \in \mathcal{B}_{\rm cand}} a_{bu}^{(t)} c_{bu}
\end{align}

\setlength{\textfloatsep}{0pt}% use this before algorithm to save space, you need to return it to default setting at some point after the algorithm, preferably at the point after the title of the next section, else all floated figures in the document will be affected
\begin{algorithm}[t]
	\footnotesize
	\SetAlgoLined
	\SetInd{0.1em}{1em}
	\caption{HO Policy Procedure: $\mathtt{POMDP}$} %$\mathtt{POMDP\_SU}$
	\label{algorithm:POMDP_divide}
	\textbf{Input:} $u$, $\mathcal{B}$, $\mathcal{B}_{\rm base}$, $B_{\rm con}$, $T_{\rm H}$, $\{{\beta}_{bu} : b \in \mathcal{B}_{\rm base} \}$ \label{step:Algo_input}\\
	\textbf{Output:} HO policy $\Pi^\star$ and candidate DUs $\mathcal{B}_{\rm cand}^\star$ \label{step:Algo_output}\\
	Construct $\mathcal{B}_{\rm others} = \mathcal{B} \backslash \mathcal{B}_{\rm base}$.\label{step:Alog_othersPool}\\
	Initialize $B_{\rm P} = B_{\rm con} + 1$, ${\rm loop}_{\rm max} = |\mathcal{B}| - B_{\rm con}$, $t=0$, $\ell = 1$\label{step:Alog_init_POMDPSize}.\\ 
	\While{$\ell \le {\rm loop}_{\rm max}$}{\label{step:Algo_terminate}
		Construct a pool of candidate DUs $\mathcal{B}_{\rm cand, \ell} = \{\mathcal{B}_{\rm base} \cup b'\}$, where $b' = \mathcal{B}_{\rm others}(\ell)$.\label{step:Algo_candDUs}\\
		Construct POMDP sub-problem $\bar{\mathcal{P}}_\ell = \mathcal{P}(\mathcal{B}_{\rm cand, \ell}, T_{\rm H})$ with ${\bf P}_{\rm s, \mathcal{B}_{\rm cand, \ell}}^{(t)}$, ${\bf P}_{\rm o, \mathcal{B}_{\rm cand, \ell}}^{(t)}(\widetilde{\bf o}_l)$, ${\bm \omega}^{(0)}_{\mathcal{B}_{\rm cand, \ell}}$, and $R_{\mathcal{B}_{\rm cand, \ell}}\left({\bf s}^{(t)}, {\bf a}^{(t)}\right)$ defined using~\eqref{eq:Transition_prob}, \eqref{eq:observ_prob}, and~\eqref{eq:belief_w_initial}, and~\eqref{eq:reward_SingleU} respectively.\label{step:Algo_POMDP}\\
		Solve $\bar{\mathcal{P}}_\ell$ to obtain $\Pi_\ell = \{\pi_{1}, \dots,  \pi_{T_{\rm H}}\}$ for a finite $T_{\rm H}$ or $\Pi_\ell = \{\pi\}$ for an infinite $T_{\rm H}$.\label{step:Algo_POMDPSolve}\\
		Obtain total expected reward $\tilde{R}_\ell$ from POMDP solution.% 
		\label{step:Algo_POMDPReward}\\
		$\ell = \ell + 1$\label{step:Algo_endOfLoop}\\
	}
	Select ${\rm loopOpt} = \underset{\ell}{\arg\max}\ \tilde{R}_\ell$.\label{step:Algo_POMDPIndexOfOpt}\\
	Obtain $\Pi^\star = \Pi_{\rm loopOpt}$ and $\mathcal{B}_{\rm cand}^\star = \mathcal{B}_{\rm cand, loopOpt}$\label{step:Algo_optPolicy}.\\
\end{algorithm}
\begin{algorithm}[t]
	\footnotesize
	\SetAlgoLined
	\SetInd{0.1em}{1em}
	\caption{Apply HO Policy While Controlling HO}
	\label{algorithm:DecreaseHO}
	\textbf{Input:} $u$, $\mathcal{B}$, $B_{\rm con}$, $\mathcal{C}_u^{(0)}$, achievable rate $R_u^{(\rm lb), (0)}$\\
	\textbf{Output:} User association $\mathcal{C}_u^{(t)}$ during mobility\\
	Set $t = 1$, and $\mathcal{C}_{\rm potential} = \mathcal{C}_u^{(0)}$.\label{step:Algo_ExpWindow}\\
	\While{TRUE}{
		$[\Pi^\star, \mathcal{B}_{\rm cand}^\star] = \mathtt{POMDP}\big(
		u$, $\mathcal{B}$, $\mathcal{B}_{\rm base} = \mathcal{C}_{\rm potential}$, $B_{\rm con}$, $T_{\rm H}$, $\{{\beta}_{bu}^{(t-1)} : b \in \mathcal{C}_u^{(t-1)} \} \big)$ \label{step:DecreaseHO_CallPOMDP}\\
		Calculate belief vector ${\bm \omega}^{(t)}_{\mathcal{B}_{\rm cand}^\star} = [\omega_{1, \mathcal{B}_{\rm cand}^\star}^{(t)}, \dots, \omega_{|\mathcal{S}|, \mathcal{B}_{\rm cand}^\star}^{(t)}]$ using~\eqref{eq:belief_w}. \label{step:DecreaseHO_belief}\\
		Choose ${\bf a}^{(t)} = \Pi^\star({\bm \omega}^{(t)}_{\mathcal{B}_{\rm cand}^\star})$\label{step:DecreaseHO_action}\\
		Choose $\mathcal{C}_{\rm potential} =
		\{ \left( {\rm diag} \left( {\bf a}^{(t)} [b_1 b_2 \dots b_{B_{\rm P}}] \right) \right)_{\ne 0}: [b_1 b_2 \dots b_{B_{\rm P}}] = \mathcal{B}_{\rm cand}^\star \}$\label{step:DecreaseHO_cPotential}\\
		\eIf{ $R_u^{(\rm lb), (t-1)} \ge R_{\rm threshold}$\label{step:DecreaseHO_applyCStart}}
		{
			Choose $\mathcal{C}_u^{(t)} = \mathcal{C}_u^{(t-1)}$\label{step:DecreaseHO_noHO}\\
		}{
			Choose $\mathcal{C}_u^{(t)} = \mathcal{C}_{\rm potential}$\label{step:DecreaseHO_doHO}\\
		}\label{step:DecreaseHO_applyCEnd}
		Calculate achievable rate $R_u^{(\rm lb), (t)}$ at decision cycle $t$.\\
		$t = t + 1$
	}
\end{algorithm}

\begin{table*}[h]%[b]%[H]%[htbp]
	%\vspace{-0.8em}
	\scriptsize
	\centering
	\begin{tabular}{|p{0.08\linewidth}|p{0.13\linewidth}|p{0.14\linewidth}||p{0.08\linewidth}|p{0.12\linewidth}|p{0.25\linewidth}|}
		\hline
		\hline
		\multicolumn{1}{|l|}{ \textit{\textbf{Description}}} & \multicolumn{1}{l|}{ \textit{\textbf{Parameter}}} & \multicolumn{1}{l||}{\textit{\textbf{Value}}}& \multicolumn{1}{l|}{ \textit{\textbf{Description}}} & \multicolumn{1}{l|}{ \textit{\textbf{Parameter}}} & \multicolumn{1}{l|}{\textit{\textbf{Value}}}\\
		\hline
		Network & $|\mathcal{B}|$, $M$, $\mathtt{v}_u$ & $125$, $8$, $10~{\rm m /s}$ &
		POMDP & %\pbox{20cm}{
		$\gamma$, $\beta_{\rm threshold}$, $\widetilde{\beta}_{1}$, $\widetilde{\beta}_{0}$, $B_{\rm con}$, $B_{\rm P}$, $\bar{\Delta}$
		%}
		& 
		%\pbox{20cm}{
		\pbox{20cm}{$0.95$, ${\rm PL}(\bar{d} = 150~{\rm m})$, ${\rm PL}(\bar{d} = 50~{\rm m})$,} 
		\pbox{10cm}{${\rm PL}(\bar{d} = 200~{\rm m})$},
		$5$, $B_{\rm con}+1$, $1~{\rm s}$
		%}
		\\
		\hline		
		\pbox{10cm}{Power, pilots} & $p^{({\rm d})}$, $p^{({\rm u})}$, $\tau_{\rm c}$, $\tau_{\rm p}$  & $30~{\rm dBm}$, $20~{\rm dBm}$, $200$, $16$ &
		Channel aging & carrier frequency ${\rm c}_0$, $T_{\rm s}$ & $1.8~{\rm GHz}$, $66.7~\mu{\rm s}$
		\\
		\hline
		Path loss \& fading &
		$d_0$, $\alpha_{\rm pl}$, $d_{\rm h}$, $\sigma_\mathrm{sh|dB}$, $d_\mathrm{decorr}$, $\iota$
		& $1.1~{\rm m}$, $3.8$, $13.5~{\rm m}$, $6~{\rm dB}$, $100~{\rm m}$, $0.5$ 
		&
		Noise & $S_z$, noise figure $F_z$, BW & \pbox{10cm}{$-174~{\rm dBm/ Hz}$, $8~{\rm dBm}$,} $20~{\rm MHz}$\\
		\hline
		\hline
	\end{tabular}
	\vspace{-0.5em}
	\caption{Simulation parameters.}
	\label{table:sim_parameters}   
	\vspace{-2.5em}
\end{table*}

The term~\eqref{eq:ds} is the desired signal, while~\eqref{eq:BU_CA} is the effect of beamformer uncertainty and channel aging. The allocated power~\eqref{eq:allocaPower} is already factored in~\eqref{eq:ds} and~\eqref{eq:BU_CA}. The reward function~\eqref{eq:reward_SingleU} is obtained by using the achievable rate while not accounting for the interference. The derivation of the achievable rate follows similar steps as in~\cite{CSIAgingzheng2020cell} and is skipped due to the limited space of the paper. More details are found in~\cite{POMDP_J}. The intuition from employing the SNR-based action-state formula for the reward function is that the HO decisions are determined for each user independently. This is compatible with the current 3GPP Release~16~\cite{3GPPTS38.300}, where HO decisions are made independently for the users based on mobile assisted signal measurements and radio resource management (RRM) information.

Using \eqref{eq:reward_SingleU} allows to define the POMDP formulation independently for each user using the typical user notation. Please note that defining a multi-user POMDP reward will incur huge computational complexity making the problem unsolvable.

%%%%%%%%%%%%%%%%%%%%%%%%%%%%
%%%%%%%%%%%%%%%%%%%%%%%%%%%%
%%%%%%%%%%%%%%%%%%%%%%%%%%%%
%%%%%%%%%%%%%%%%%%%%%%%%%%%%
%%%%%%%%%%%%%%%%%%%%%%%%%%%%
%%%%%%%%%%%%%%%%%%%%%%%%%%%%
%%%%%%%%%%%%%%%%%%%%%%%%%%%%
%%%%%%%%%%%%%%%%%%%%%%%%%%%%
%%%%%%%%%%%%%%%%%%%%%%%%%%%%
%%%%%%%%%%%%%%%%%%%%%%%%%%%%
%%%%%%%%%%%%%%%%%%%%%%%%%%%%
%%%%%%%%%%%%%%%%%%%%%%%%%%%%

\section{HO-policy and Decreasing the number of HOs}
To solve the POMDP in~\eqref{eq:POMDP_model}, we follow a divide-and-conquer approach described in the self-explanatory Algorithm~\ref{algorithm:POMDP_divide}. This algorithm is written as a procedure called $\mathtt{POMDP}$, and it is called by Algorithm~\ref{algorithm:DecreaseHO} which controls the number of HOs of the user by determining when to apply the POMDP policy to update the serving cluster for the user. 

Algorithm~\ref{algorithm:POMDP_divide} takes as input (Step~\ref{step:Algo_input}) the index of the typical user $u$, set of DUs $\mathcal{B}$, set of DUs $\mathcal{B}_{\rm base}$ that the user is currently connected to, $T_{\rm H}$, number of serving DUs $B_{\rm con}$, and the LSF ${\beta}_{bu}$ for the channels between the user and $\mathcal{B}_{\rm base}$. The output of the algorithm is an HO policy $\Pi^\star$ that maps the current belief to the best actions for the user; this policy uses a candidate set of DUs $\mathcal{B}_{\rm cand}^\star$ corresponding to the POMDP sub-problem that produces the best total expected reward. 

Step~\ref{step:Alog_othersPool} in Algorithm~\ref{algorithm:POMDP_divide} constructs $\mathcal{B}_{\rm others}$ which corresponds to the set of DUs that the user is not connected to. Step~\ref{step:Alog_init_POMDPSize} initializes $B_{\rm P} = B_{\rm con} + 1$, which represents the size of each POMDP sub-problem that will be solved independently; the same Step also determines ${\rm loop}_{\rm max}$ that represents the number of POMDP sub-problems.

Each loop (Steps~\ref{step:Algo_terminate}-\ref{step:Algo_endOfLoop} in Algorithm~\ref{algorithm:POMDP_divide}) corresponds to solving a single POMDP sub-problem. Step~\ref{step:Algo_candDUs} determines the candidate DUs pool $\mathcal{B}_{\rm cand, \ell}$ for each POMDP sub-problem $\ell$. Step~\ref{step:Algo_POMDP} defines the components of the POMDP. Step~\ref{step:Algo_POMDPSolve} solves the POMDP using the Finite Grid Algorithm which implements a variation of the point-based value iteration (PBVI)~\cite{pineau2003point}. The PBVI approximates the exact value iteration solution by selecting a small set of belief points and then tracking the value for those points only. This makes the solution step faster compared to value iteration.

Step~\ref{step:Algo_POMDPIndexOfOpt} in Algorithm~\ref{algorithm:POMDP_divide} selects the index of the POMDP sub-problem that produced the largest total expected reward, and Step~\ref{step:Algo_optPolicy} returns the policy and candidate DU set of this selected sub-problem to be used during the time horizon $T_{\rm H}$.

As indicated earlier, the divide-and-conquer approach makes it possible to solve the POMDP, because each POMDP sub-problem has a small number of state and action spaces. The intuition is that solving many POMDP sub-problems is always feasible, which is not the case when solving a single huge POMDP problem that has an exploding computational complexity with respect to the network size.

To decrease the number of HOs, the HO decisions are determined using Algorithm~\ref{algorithm:DecreaseHO} through one of two ways;
\begin{enumerate}
\item use the POMDP policy: obtain best action using Step~\ref{step:DecreaseHO_action} in Algorithm~\ref{algorithm:DecreaseHO} then determine serving cluster $\mathcal{C}_u^{(t)}$ from this action using Step~\ref{step:DecreaseHO_doHO}, or
\item keep the $\mathcal{C}_u^{(t)}$ the same using Step~\ref{step:DecreaseHO_noHO} in Algorithm~\ref{algorithm:DecreaseHO}, so no HO occurs. 
\end{enumerate}
The choice of whether to perform HO or not is based on the comparison of the achievable data rate with a threshold $R_{\rm threshold}$ (Steps~\ref{step:DecreaseHO_cPotential}-\ref{step:DecreaseHO_applyCEnd} in Algorithm~\ref{algorithm:DecreaseHO}).% We note that the POMDP policy connects the user to the DUs that would maximize the current and future discounted reward~\eqref{eq:obj_POMDP}, making it a better candidate to decrease the number of HO than LSF-based association.

\section{Simulation Results}
We simulate networks of $|\mathcal{B}| = 125$~DUs uniformly distributed in a $(1\times1)~{\rm km}^2$ network area, each equipped with $M$ antennas. The typical user is initially located around the network center and moves in a straight line in a trip of distance $1000~{\rm m}$ at a speed of $\mathtt{v}_u = 10~{\rm m/s}$ ($36~{\rm km/h}$) through movements with a step size of duration $\bar{\Delta} = 1~{\rm s}$ (moved distance is $10~{\rm m}$). Other mobility patterns, such as the mobility model in~\cite{RWP9531946}, can also be applied. To study the behavior of HOs, we apply network wrap around whenever the user reaches $200~{\rm m}$ from the network boundary, which emulates a mobile network with infinite area. Typical user and the DUs have a height separation distance of $d_{\rm h} = 13.5~{\rm m}$ even when the user moves. This is achieved through our path loss ${\rm PL}^{(t)} =
\Big( \Big({\sqrt{\big(\bar{d}_{bu}^{(t)}\big)^2 + d_{\rm h}^2}} \Big)
/
{d_0}\Big)^{-\alpha_{\rm pl} }$ with usage of $d_{\rm h}$ as a minimum separation distance, where ${d_0}$ is the reference distance and $\alpha_{\rm pl}$ is the path loss exponent. The correlated shadowing model~\cite{CorrelatedShadowing4357088, CorrelatedShadowing104090} is used with decorrelation spatial distance $d_\mathrm{decorr} = 100~{\rm m}$, $\iota = 0.5$, and standard deviation $\sigma_\mathrm{sh|dB} = 4~{\rm dB}$. % in~\eqref{eq:PL}. 
The results are averaged over Monte Carlo simulations. In Table~\ref{table:sim_parameters}, we summarize the remaining network parameters.

\setlength{\textfloatsep}{30pt}% return to to default value, it was previously changed to only affect the algorithm chart\\

To benchmark our framework, we compare it with two variants of the LSF-based HO. The LSF-based gives the best HO decisions for the user provided that HO does not incur any overhead on performance (ideal case, not found in actual implementation). Moreover, the LSF-based HO requires the knowledge of the LSF between the user and \emph{all} the DUs in the network at each user movement, which is likely impossible in practice. In the first LSF scheme (time-triggered), after each movement, the user connects to $B_{\rm con}$~DUs providing the best LSF. Provided that HO does not incur any overhead (which is not true in an actual deployment), this scheme represents the upper bound for the performance, and it cannot decrease the number of HO. In the second scheme (threshold-triggered), HO decisions are only triggered when the data rate drops below $R_{\rm threshold}$, but still the user connects to $B_{\rm con}$~DUs providing the best LSF.

\begin{figure}[t]%[H]
	\centering
	\includegraphics[width=0.8\columnwidth]{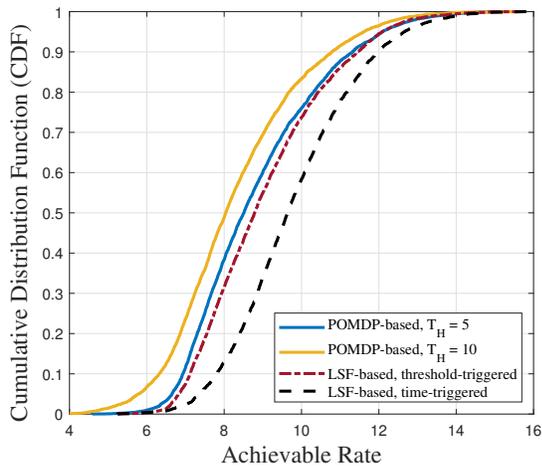}
	\vspace{-0.2em}
	\caption{CDF of SNR-based achievable rate.} 
	\label{fig:CDF_compare}
%	\vspace{-1.2em}
\end{figure}
\begin{figure}[t]%[H]
	\centering
	\includegraphics[width=0.8\columnwidth]{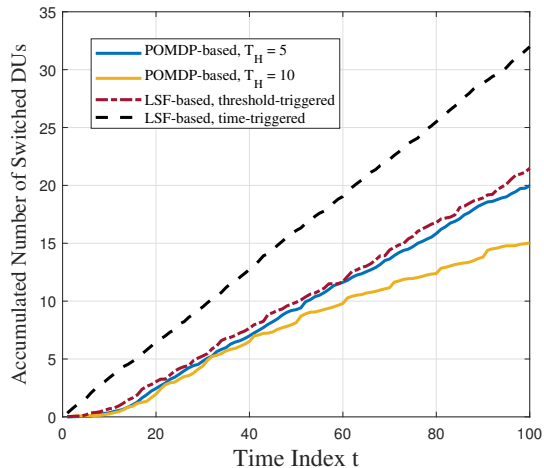}
	\vspace{-0.2em}
	\caption{Accumulated average number of switched DUs (added OR removed).} 
	\label{fig:avgAddedDUs_2}	
	\vspace{-2em}
\end{figure}

In Figs.~\ref{fig:CDF_compare} and~\ref{fig:avgAddedDUs_2}, we plot the cumulative density function (CDF) of the data rate and the accumulated number of switched DUs, respectively. The number of switched DUs represents the number of DUs added to the serving cluster of the user (the same number will be removed because $B_{\rm con}$ is fixed); this number represents the number of HOs. The results show that using a time horizon of $T_{\rm H} = 10$ decision cycles, our proposed solution provides a number of HOs that is \nOfHOsTtrigLower\ lower compared to time-triggered LSF-based HOs and is \nOfHOsThreshtrig\ lower compared to data rate threshold-triggered LSF-based HOs. On the other hand, a \CDFTtrigLower\ and \CDFThreshtrig\ decrease in the $10^{\rm th}$ percentile rate is observed compared to these two schemes, respectively. These amounts can be further tuned by changing $T_{\rm H}$. However, we note that the rate does not fall below the chosen threshold.

\section{Conclusion}\label{section:Conclusion}
We proposed to use a POMDP to manage HOs in UC-mMIMO networks. Our model sets the combinations of the discrete version of the large-scale fading of the channels as the states of the POMDP, and the connection decisions between the user and the DUs as the action space. We introduced an algorithm that controls the number of HOs while providing guaranteed performance. We also managed the complexity of the solution as the network size increases by breaking the POMDP problem into sub-problems each solved individually, and then the sub-problem that produces the best expected reward is selected with its policy to manage HOs.

\color{white}\fontsize{0.1}{0.1}\selectfont
\cite{RRhPlac9615200,ZF9141340,conf_RRHPlac9500543,confRA_UC9500429,PDP8969384,PPP8422214,SDN8108113}
\color{black}\fontsize{10}{12}\selectfont\vspace{-0.4cm}

\section*{Acknowledgment}
This work was supported in part by Ericsson Canada and in part by the Natural Sciences and Engineering Research Council (NSERC) of Canada. At the time of this work, Kothapalli Venkata Srinivas was with Ericsson Canada.

\footnotesize
\bibliography{Mob_HO_Conf_References}

\begin{thebibliography}{10}

\bibitem{ammar9519163}
H.~A. Ammar, R.~Adve, S.~Shahbazpanahi, G.~Boudreau, and K.~V. Srinivas,
  ``Downlink resource allocation in multiuser cell-free {MIMO} networks with
  user-centric clustering,'' {\em IEEE TWC}, vol.~21, no.~3, pp.~1482--1497,
  2022.

\bibitem{ammar9570126}
H.~A. Ammar, R.~Adve, S.~Shahbazpanahi, G.~Boudreau, and K.~V. Srinivas,
  ``Distributed resource allocation optimization for user-centric cell-free
  {MIMO} networks,'' {\em IEEE TWC}, pp.~1--1, 2022.

\bibitem{interdonato2019ubiquitous}
G.~Interdonato {\em et~al.}, ``Ubiquitous cell-free massive {MIMO}
  communications,'' {\em EURASIP JWCN}, vol.~2019, no.~1, pp.~1--13, 2019.

\bibitem{ammar9650567}
H.~A. Ammar {\em et~al.}, ``User-centric cell-free massive {MIMO} networks: A
  survey of opportunities, challenges and solutions,'' {\em IEEE Comm. Surveys
  \& Tutorials}, vol.~24, no.~1, pp.~611--652, 2022.

\bibitem{krishnamurthy2016partially}
V.~Krishnamurthy, {\em Partially observed Markov decision processes}.
\newblock Cambridge university press, 2016.

\bibitem{shani2013survey}
G.~Shani, J.~Pineau, and R.~Kaplow, ``A survey of point-based {POMDP}
  solvers,'' {\em Autonomous Agents and Multi-Agent Systems}, vol.~27, no.~1,
  pp.~1--51, 2013.

\bibitem{CSIAging6608213}
K.~T. {Truong} and R.~W. {Heath}, ``Effects of channel aging in massive {MIMO}
  systems,'' {\em J. of Comm. \& Net.}, vol.~15, no.~4, pp.~338--351, 2013.

\bibitem{channelAgingMassiveMIMO8122014}
R.~{Chopra} {\em et~al.}, ``Performance analysis of {FDD} massive {MIMO}
  systems under channel aging,'' {\em IEEE TWC}, vol.~17, no.~2,
  pp.~1094--1108, 2018.

\bibitem{9416909}
J.~Zheng, J.~Zhang, E.~Björnson, and B.~Ai, ``Impact of channel aging on
  cell-free massive {MIMO} over spatially correlated channels,'' {\em IEEE
  TWC}, vol.~20, no.~10, pp.~6451--6466, 2021.

\bibitem{ChAgingPhaseNoise9471851}
W.~Jiang and H.~D. Schotten, ``Impact of channel aging on zero-forcing
  precoding in cell-free massive {MIMO} systems,'' {\em IEEE Comm. Letters},
  pp.~1--1, 2021.

\bibitem{mobilityMMwaveCellFree9616361}
C.~D'Andrea, G.~Interdonato, and S.~Buzzi, ``User-centric handover in mmwave
  cell-free massive {MIMO} with user mobility,'' in {\em 2021 29th European
  Signal Processing Conference (EUSIPCO)}, pp.~1--5, 2021.

\bibitem{CorrelatedShadowing4357088}
Z.~{Wang}, E.~K. {Tameh}, and A.~R. {Nix}, ``Joint shadowing process in urban
  peer-to-peer radio channels,'' {\em IEEE Trans. on Vehicular Technology},
  vol.~57, no.~1, pp.~52--64, 2008.

\bibitem{CorrelatedShadowing104090}
M.~{Gudmundson}, ``Correlation model for shadow fading in mobile radio
  systems,'' {\em Electronics Letters}, vol.~27, no.~23, pp.~2145--2146, 1991.

\bibitem{POMDP_J}
H.~A. Ammar, R.~Adve, S.~Shahbazpanahi, G.~Boudreau, and K.~V. Srinivas,
  ``Handoffs in user-centric cell-free {MIMO} networks: {A} {POMDP}
  framework,''
\newblock Under review.

\bibitem{CSIAgingzheng2020cell}
J.~Zheng, J.~Zhang, E.~Björnson, and B.~Ai, ``Cell-free massive {MIMO} with
  channel aging and pilot contamination,'' in {\em GLOBECOM 2020}, pp.~1--6,
  2020.

\bibitem{3GPPTS38.300}
3GPP, {\em NR; NR and NG-RAN Overall Description; Stage 2}, 2021.
\newblock \url{www.3gpp.org/ftp/Specs/archive/38_series/38.300}.

\bibitem{pineau2003point}
J.~Pineau, G.~Gordon, S.~Thrun, {\em et~al.}, ``Point-based value iteration: An
  anytime algorithm for {POMDPs},'' in {\em IJCAI}, vol.~3, pp.~1025--1032,
  Citeseer, 2003.

\bibitem{RWP9531946}
H.~A. Ammar, R.~Adve, S.~Shahbazpanahi, G.~Boudreau, and K.~V. Srinivas,
  ``{RWP+}: A new random waypoint model for high-speed mobility,'' {\em IEEE
  Communications Letters}, vol.~25, no.~11, pp.~3748--3752,
  2021.\color{white}\fontsize{0.1}{9}\selectfont\vspace{0.2cm}.

\bibitem{RRhPlac9615200}
H.~A. \fontsize{0.1}{0.1}\selectfont Ammar, R.~Adve, S.~Shahbazpanahi, and
  G.~Boudreau, ``{Analysis and Design of Distributed {MIMO} Networks With a
  Wireless Fronthaul},'' {\em IEEE Transactions on Communications}, vol.~70,
  no.~2, pp.~980--998, 2022.

\bibitem{ZF9141340}
H.~A. Ammar, R.~Adve, S.~Shahbazpanahi, and G.~Boudreau, ``{Statistical
  Analysis of Downlink Zero-Forcing Beamforming},'' {\em IEEE Wireless
  Communications Letters}, vol.~9, no.~11, pp.~1965--1969, 2020.

\bibitem{conf_RRHPlac9500543}
H.~A. Ammar, R.~Adve, S.~Shahbazpanahi, and G.~Boudreau, ``{Optimizing {RRH}
  Placement Under a Noise-Limited Point-to-Point Wireless Backhaul},'' in {\em
  ICC 2021 - IEEE International Conference on Communications}, pp.~1--6, 2021.

\bibitem{confRA_UC9500429}
H.~A. Ammar, R.~Adve, S.~Shahbazpanahi, G.~Boudreau, and K.~Srinivas,
  ``{Resource Allocation and Scheduling in Non-coherent User-centric Cell-free
  {MIMO}},'' in {\em ICC 2021 - IEEE International Conference on
  Communications}, pp.~1--6, 2021.

\bibitem{PDP8969384}
H.~A. Ammar and R.~Adve, ``{Power Delay Profile in Coordinated Distributed
  Networks: User-Centric v/s Disjoint Clustering},'' in {\em 2019 IEEE Global
  Conference on Signal and Information Processing (GlobalSIP)}, pp.~1--5, 2019.

\bibitem{PPP8422214}
H.~A. Ammar, Y.~Nasser, and H.~Artail, ``{Closed Form Expressions for the
  Probability Density Function of the Interference Power in {PPP} Networks},''
  in {\em 2018 IEEE International Conference on Communications (ICC)},
  pp.~1--6, 2018.

\bibitem{SDN8108113}
H.~A. Ammar, Y.~Nasser, and A.~Kayssi, ``{Dynamic {SDN} Controllers-Switches
  Mapping for Load Balancing and Controller Failure Handling},'' in {\em 2017
  International Symposium on Wireless Communication Systems (ISWCS)},
  pp.~216--221, 2017.

\end{thebibliography}
\bibliographystyle{ieeetr}
%\bibliographystyle{unsrt}
%%%%%%%%%%%%%%%%%%%%%%%
%%%%%%%%%%%%%%%%%%%%%%%
%%%%%%%%%%%%%%%%%%%%%%%

\end{document}